\documentclass[aps,prl,twocolumn,showpacs,superscriptaddress,floatfix]{revtex4-1}
\usepackage{graphicx}
\usepackage{dcolumn}
\usepackage{bm}
\usepackage{amssymb}
\begin{document}

\title{From Spin Glass to Quantum Spin Liquid Ground States in Molybdate Pyrochlores}
\author{L.Clark}
\affiliation{Department of Physics and Astronomy, McMaster University, Hamilton, Ontario, L8S 4M1, Canada}
\author{G. J. Nilsen}
\affiliation{Institute Laue-Langevin, 6 Rue Jules Horowitz, 38042 Grenoble, France}
\author{E. Kermarrec}
\affiliation{Department of Physics and Astronomy, McMaster University, Hamilton, Ontario, L8S 4M1, Canada}
\author{G. Ehlers}
\affiliation{Neutron Scattering Science Division, Oak Ridge National Laboratory, Oak Ridge, Tennessee 37831, USA}
\author{K. S. Knight}
\affiliation{ISIS Facility, Rutherford Appleton Laboratory, Didcot, Oxfordshire, OX11 0QX, UK.}
\author{A. Harrison}
\affiliation{Diamond Light Source,  Harwell Science and Innovation Campus, Didcot, Oxfordshire, OX11 0QX, UK}
\author{J. P. Attfield}
\affiliation{CSEC and School of Chemistry, University of Edinburgh, Mayfield Road, Edinburgh, EH9 3JZ, UK}
\author{B. D. Gaulin}
\affiliation{Department of Physics and Astronomy, McMaster University, Hamilton, Ontario, L8S 4M1, Canada}
\affiliation{Brockhouse Institute for Materials Research, Hamilton, Ontario, L8S 4M1, Canada}
\affiliation{Canadian Institute for Advanced Research, 180 Dundas St. W., Toronto, Ontario, M5G 1Z8, Canada}
\date{\today}

\begin{abstract}
We present new magnetic heat capacity and neutron scattering results for two magnetically frustrated molybdate pyrochlores: $S=1$ oxide Lu$_2$Mo$_2$O$_7$ and $S={\frac{1}{2}}$ oxynitride Lu$_2$Mo$_2$O$_5$N$_2$. Lu$_2$Mo$_2$O$_7$ undergoes a transition to an unconventional spin glass ground state at $T_f~{\sim}~16$ K. However, the preparation of the corresponding oxynitride tunes the nature of the ground state from spin glass to quantum spin liquid. The comparison of the static and dynamic spin correlations within the oxide and oxynitride phases presented here reveals  the crucial role played by quantum fluctuations in the selection of a ground state. Furthermore, we estimate an upper limit for a gap in the spin excitation spectrum of the quantum spin liquid state of the oxynitride of ${\Delta}~{\sim}~0.05$ meV or ${\frac{\Delta}{|\theta|}}\sim0.004$, in units of its antiferromagnetic Weiss constant ${\theta}~{\sim}-121$ K.
\end{abstract}

\pacs{75.10.Kt, 75.40.Cx, 75.50.Lk} 

\maketitle

Geometric magnetic frustration arises when the arrangement of magnetic moments on an ordered lattice prevents the simultaneous satisfaction of each pair-wise magnetic exchange interaction {\cite{Greedan2001}}. This can generate a macroscopic ground state degeneracy, whereby the the spins on the vertices of the  frustrated lattice fluctuate within a manifold of non-ordered states. In materials that possess both strong geometric frustration and quantum ($S={\frac{1}{2}}$) magnetic moments, quantum fluctuations can induce a quantum spin liquid (QSL) ground state {\cite{Anderson1973}}, {\cite{Balents2010}} and allow the system to remain dynamic and avoid any symmetry breaking phase transition, even in the thermodynamic limit of $T=0$ K. There are now several promising experimental realisations of materials that host this exotic QSL state in two-dimensional frustrated lattices {\cite{HerbertsmithiteMuSR}}, {\cite{deVries2009}}, {\cite{Kapellasite2012}}, {\cite{Clark_DQVOF_2013}}, {\cite{YamashitaScience2010}}. However, examples of three-dimensional QSL candidates, such as a $S={\frac{1}{2}}$ pyrochlore antiferromagnet that possesses a frustrated magnetic lattice of corner sharing tetrahedra, remain scarce {\cite {RossPRX}}. There is some uncertainty regarding the exact nature of the spin liquid ground state of a $S={\frac{1}{2}}$ pyrochlore antiferromagnet {\cite{Canals1998}}, {\cite{Berg2003}}, {\cite{Hermele2004}}  and the synthesis and study of such new materials remain important tasks {\cite{Harrison2004}}.

Magnetic frustration in the rare earth molybdate pyrochlores, $R_2$Mo$_2$O$_7$, arises from exchange interactions between Mo$^{4+}$ $4d^1$ $S=1$ spins residing on a network of vertex sharing tetrahedra. Among the phenomena observed in these materials {\cite{ PhysRevLett_MIT}}, the spin glass state in the apparently disorder free Y$_2$Mo$_2$O$_7$ is of particular interest {\cite{Greedan1986}}, {\cite{PhysRevLettGingra1997}}, {\cite{GardnerPRL1999}}, {\cite{JPPS2000}}, {\cite{SilversteinPRB2014}}. As this state - a spin glass in the absence of any significant chemical or bond disorder - is very unusual, its realisation in a similarly simple material may be revealing. Lu$_2$Mo$_2$O$_7$ is a sister compound to Y$_2$Mo$_2$O$_7$ and exhibits strong antiferromagnetic exchange, with the Weiss constant ${\theta}=-160$ K, as well as a spin freezing transition at $T_f~{\sim}~16$ K {\cite{ClarkJSSC2013}}. These parameters are remarkably close to the ${\theta}~{\sim}~-200$ K and $T_f~{\sim}~22$ K displayed by Y$_2$Mo$_2$O$_7$.

A group of materials related to the $R_2$Mo$_2$O$_7$ pyrochlores are the corresponding oxynitride phases. Oxynitrides are mixed anion materials that are often prepared by a direct nitridation of an oxide precursor. In the case of the molybdate pyrochlores, this means that the cubic pyrochlore structure is retained. The incorporation of the nitride (N$^{3-}$) anion into the oxide (O$^{2-}$) framework is accompanied by oxidation of the molybdenum cations to maintain overall charge neutrality. Oxynitrides of composition $R_2$Mo$_2$O$_5$N$_2$ possess $S={\frac{1}{2}}$ spins due to the Mo$^{5+}$ $4d^1$ oxidation state and are thus excellent candidates for the study of QSL phenomena {\cite{YangChemMater2010}}. Here we present a comparison of Lu$_2$Mo$_2$O$_7$ and Lu$_2$Mo$_2$O$_5$N$_2$ by way of low temperature heat capacity and neutron scattering studies.

Polycrystalline Lu$_2$Mo$_2$O$_7$ was synthesised as described elsewhere {\cite{ClarkJSSC2013}}. An oxynitride phase with a composition close to the ideal $S={\frac{1}{2}}$ Lu$_2$Mo$_2$O$_5$N$_2$ stoichiometry was prepared by thermal ammonolysis of a Lu$_2$Mo$_2$O$_7$ precursor. As shown in the Supplemental Material (SM), Rietveld refinement of the $Fd{\bar{3}m}$ cubic pyrochlore model to powder neutron diffraction data confirms that the oxynitride adopts this structure with oxide and nitride anions disordered over the two available anion sites. Magnetic susceptibility data for the oxynitride (see SM) reveal an absence of long range order or spin freezing transitions down to $2$ K despite the large antiferromagnetic ${\theta}=-121(1)$ K. The small effective magnetic moment ${\mu}_{eff}=1.11(1)~{\mu}_B$ per Mo cation reflects the oxidation of Mo$^{4+}$ ($S=1$) to Mo$^{5+}$ ($S={\frac{1}{2}}$). The $64~{\%}$ reduction of the observed ${\mu}_{eff}$ from the expected spin-only value for $S={\frac{1}{2}}$ is consistent with the $67~{\%}$ reduction from the expected $S=1$ spin-only moment displayed by Lu$_2$Mo$_2$O$_7$ {\cite{ClarkJSSC2013}}. This suggests that the spin-orbit coupling interaction is significant in both of these $4d$ systems. 

\begin{figure}
\includegraphics[width=3.5in]{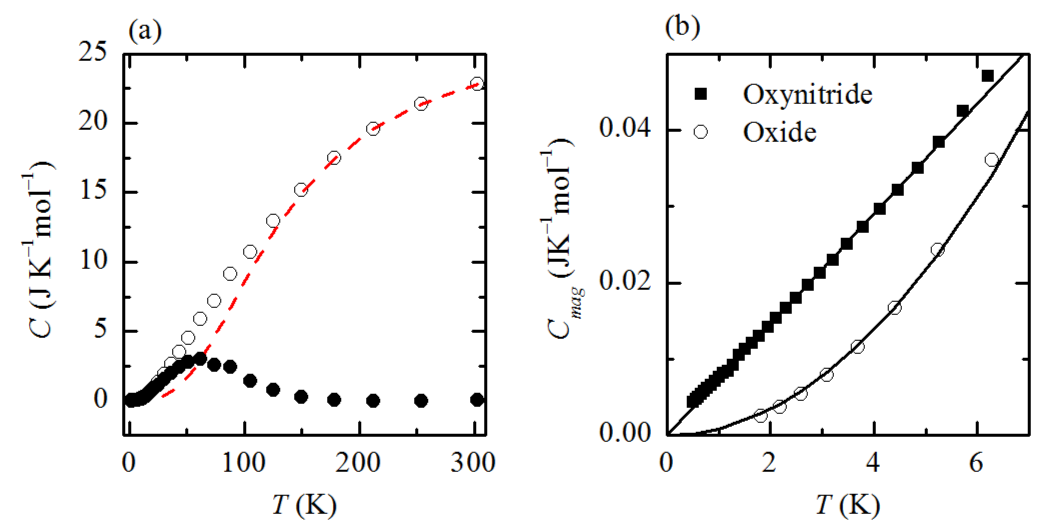}
\caption{(a) The total heat capacity $C$ of Lu$_2$Mo$_2$O$_7$ as a function of temperature (open circles) with an estimation of the lattice contribution from a Debye fit to the high temperature data (dashed line). The magnetic heat capacity $C_{mag}$ (closed circles) was obtained by substraction of this fit from the measured total. (b) The magnetic heat capacities of the oxide and oxynitride phases follow strikingly different behaviour at low temperatures.}
\label{Figure1}
\end{figure}

The heat capacity of Lu$_2$Mo$_2$O$_7$ was measured on a $9.0$ mg pressed powder pellet in a Quantum Design PPMS. In an attempt to estimate the lattice contribution to the total heat capacity, the high temperature data were modelled by the Debye equation. The fit to the data, which gave a Debye temperature ${\theta}_D=540$ K, is shown in Fig.{\ref{Figure1}}(a). Upon subtraction of the estimated lattice contribution, a broad hump centred ${{\sim}~50}$ K is observed in the magnetic heat capacity $C_{mag}$, typical of a spin glass system. The heat capacity of the oxynitride was measured on a $8.9$ mg sample over the range $0.5~-~30$ K using a $^3$He insert. Given the similar structure and formula weight of the oxide and oxynitride phases, the lattice contribution estimated for Lu$_2$Mo$_2$O$_7$ was also used to extract the magnetic heat capacity of the oxynitride. A comparison of the low temperature magnetic heat capacities of the oxide and oxynitride are shown in Fig.{\ref{Figure1}}(b). The temperature dependencies are markedly different with $C_{mag}~{\propto}~T^2$ for the oxide and $C_{mag}~{\propto}~T$ for the oxynitride.       

\begin{figure}
\includegraphics[width=3.5in]{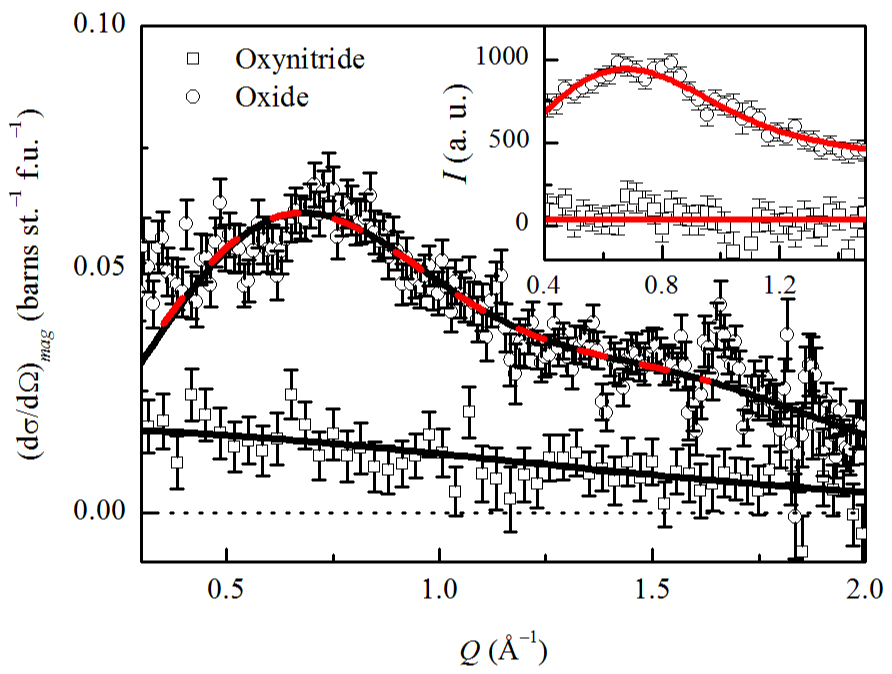}
\caption{The magnetic scattering cross sections of the oxide and oxynitride obtained from $xyz$ polarization analysis of data measured on the D$7$ spectrometer at $1.5$ K. The solid lines are fits to the data described in the text. For comparison, the inset shows the $Q$-dependence of elastic magnetic scattering measured on CNCS at the same temperature.  A scaled version of the spin correlation model fitted to the CNCS oxide data is shown on top of the D$7$ magnetic scattering cross section (dashed line).}
\label{Figure2}
\end{figure}

Diffuse magnetic neutron diffraction was measured for the oxide and oxynitride pyrochlores on the D$7$ Diffuse Scattering Spectrometer at the Institut Laue-Langevin, France {\cite{Stewart_D7}}. $xyz$ polarisation analysis was used separate the components of total neutron scattering {\cite{Scharpf1993}}, {\cite{RevSciInst2013}}. In this way, the magnetic diffuse scattering that arises from disordered spin systems can be isolated from the strong nuclear Bragg scattering that dominates a typical neutron diffraction pattern {\cite{Cywinski1999}}. Data were collected for both samples with an incident neutron wavelength ${\lambda}=4.8$ {\AA}, at which final neutron energies are integrated up to $E~{\sim}~3.5$ meV. A detailed experimental account, including a description of the absorption correction applied to the data is given in the SM. Fig.{\ref{Figure2}} shows the magnetic scattering cross sections ($d{\sigma}$/$d{\Omega}$)$_{mag}$ of the oxide and oxynitride at $1.5$ K, well below $T_f~{\sim}~16$ K. The magnetic diffuse scattering from the oxide displays a broad peak centred around $0.6$ {\AA}$^{-1}$ that indicates the presence of static short ranged molybdenum spin correlations. The data were modelled according to the expression {\cite{NilsenPRB2011}},

\begin{eqnarray}
{\Bigg(}{\frac{d{\sigma}}{d{\Omega}}}{\Bigg)}_{mag} = {\frac{2}{3}}({\gamma_n}r_0)^2{\Bigg(}{\frac{1}{2}}gF(Q){\Bigg)}^2 \nonumber \\
{\times}{\Bigg(}S(S+1)+{\sum_i}Z_i{\langle}{\mathbf{S}_0}{\cdot}{\mathbf{S}_i}{\rangle}{\frac{{\sin}Qr_i}{Qr_i}}{\Bigg)}
\end{eqnarray}
where ${\langle}{\mathbf{S}_0}{\cdot}{\mathbf{S}_i}{\rangle}$ gives the correlation between a spin and its $Z_i$ nearest neighbours at a distance $r_i$. ${\gamma}_n$ is the neutron gyromagnetic ratio, $r_0$ is the classical radius of the electron, $g$ is the $g$-factor and $F(Q)$ is the molybdenum form factor {\cite{Cryst_C}}. The best fit to the data, shown as a black solid line in Fig.{\ref{Figure2}}, was obtained by allowing for nearest- ($r_1=3.581$ {\AA}, $Z_1=6$) and next-nearest-neighbour ($r_2=6.203$ {\AA}, $Z_2=12$) correlations ${\langle}{\mathbf{S}_0}{\cdot}{\mathbf{S}_1}{\rangle}=-0.029(6)$ and ${\langle}{\mathbf{S}_0}{\cdot}{\mathbf{S}_2}{\rangle}=-0.056(7)$, respectively. Adding further exchange correlations did not improve the quality of the fit significantly. The negative sign of the spin correlations confirms their antiferromagnetic nature. In contrast, the magnetic diffuse scattering of the oxynitride at $1.5$ K is much weaker than that of the oxide and appears to follow $F(Q)^2$. These data were thus modelled using Equation $1$ with ${\langle}{\mathbf{S}_0}{\cdot}{\mathbf{S}_i}{\rangle}=0$ for all $i$ {\cite{WillsPRB2001}}. The fit is shown as the solid line against the oxynitride data in Fig.{\ref{Figure2}}. The effective magnetic moment extracted from the fit to the data, $0.11(1)~{\mu}_B$, corresponds to only $6~{\%}$ of the expected $S={\frac{1}{2}}$ spin-only value, suggesting that most of the scattering from the oxynitride is inelastic and thus outside the energy range over which D$7$ integrates scattered neutron energies {\cite{Cywinski1999}}.

In order to probe their full static and dynamic behaviour, both the oxide and oxynitride pyrochlores were studied on the Cold Neutron Chopper Spectrometer (CNCS) {\cite{Ehlers_Rev_Sci_Instrum_2011}} at the Spallation Neutron Source of the Oak Ridge National Laboratory. Measurements on both samples were performed on cooling to $1.5$ K with an incident neutron energy $E_i=3.3$ meV. The inset of Fig.{\ref{Figure2}} shows the $Q$-dependence of the elastic scattering from the oxide and oxynitride at $1.5$ K, obtained by integrating the inelastic spectra over the energy of the elastic line, $E=[-0.1,0.1]$ meV.  A broad peak at low $Q$ is observed for the oxide, which can again be modelled by Equation $1$. To confirm the consistency of the analyses of the CNCS and D$7$ data sets, a scaled version of the fit to the CNCS data is plotted with the D$7$ fit in the main panel of Fig.{\ref{Figure2}}; agreement is excellent. Remarkably, the scattering collected for the oxynitride at $1.5$ K within the narrowly defined elastic window on CNCS is consistent with {\it no elastic magnetic scattering}, as shown in the inset to Fig.{\ref{Figure2}}.

\begin{figure*}
\includegraphics[width=7in]{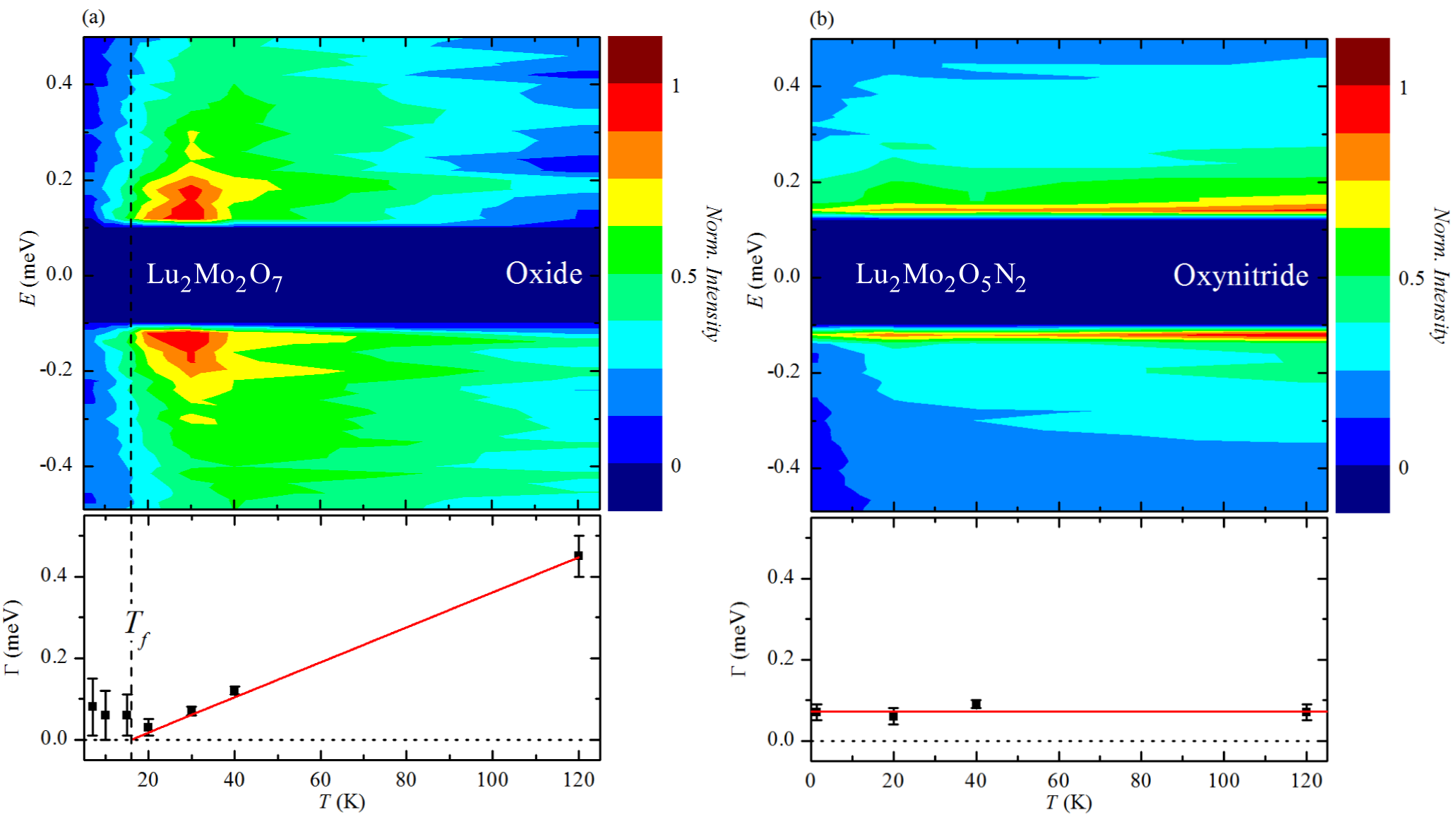}
\caption{The temperature dependence of background subtracted, normalised inelastic neutron scattering intensity integrated over $Q=[0.5,2.0]$ {\AA}$^{-1}$ (top) and the frequency width of the inelastic spectra, $\Gamma$, (bottom) for the (a) oxide and (b) oxynitride, respectively.}
\label{Figure3}
\end{figure*}

The temperature dependencies of the background corrected, normalised and $Q$-integrated ($Q=[0.5,2.0]$ {\AA}$^{-1}$) inelastic scattering for the oxide and oxynitride are shown in the top panels of Figs.{\ref{Figure3}}(a) and {\ref{Figure3}}(b), respectively. The oxide (Fig.{\ref{Figure3}}(a)) presents a rather broad frequency spectrum at higher temperatures, $T$ {\textgreater} $50$ K, well above the spin freezing transition. Upon cooling, low energy spin fluctuations develop and sharpen in frequency approaching the spin freezing transition. The transition is clearly marked by a collapse of inelastic scattering intensity into the elastic line as short range spin correlations build up and give rise to elastic intensity in the diffuse magnetic scattering peak at $0.6$ {\AA}$^{-1}$. In order to obtain a more quantitative understanding of the inelastic scattering data, energy cuts were fitted to the general form of the scattering function {\cite{GardnerPRL1999}},

\begin{equation}
S(E) = {\frac{1}{\pi}}{\chi}^{''}(E)[1+n(E)]
\end{equation}
where $n(E)$ is the Bose-Einstein thermal population factor and ${\chi}^{''}$ is the imaginary part of the dynamic susceptibility. A good phenomenological description of the data was obtained with,

\begin{equation}
{\chi}^{''}(E) = {\chi_0}{\arctan}{\Bigg(}{\frac{E}{\Gamma}}{\Bigg)}
\end{equation}
where ${\Gamma}$ is related to the frequency width of the inelastic spectrum. Fits to the data are shown in the SM. The bottom panel of Fig.{\ref{Figure3}}(a) shows the temperature dependence of ${\Gamma}$ for the oxide, which goes to zero at or near $T_f~{\sim}~16$ K, consistent with spin glass freezing. The overall dynamic behavior of  Lu$_2$Mo$_2$O$_7$ is thus very similar to that of its sister ``disorder free" spin glass pyrochlore Y$_2$Mo$_2$O$_7$  {\cite{GardnerPRL1999}}.  Again in rather stark contrast, the oxynitride displays a temperature-independent energy width of ${\Gamma}~{\sim}~0.08$ meV, consistent with that of Lu$_2$Mo$_2$O$_7$ near 30 K, almost twice $T_f$ (Fig.{\ref{Figure3}}(b)).  Therefore the temperature dependence observed in $S(E)$ for the oxynitride originates entirely from the Bose-Einstein factor in Equation $2$.

\begin{figure}
\includegraphics[width=3.5in]{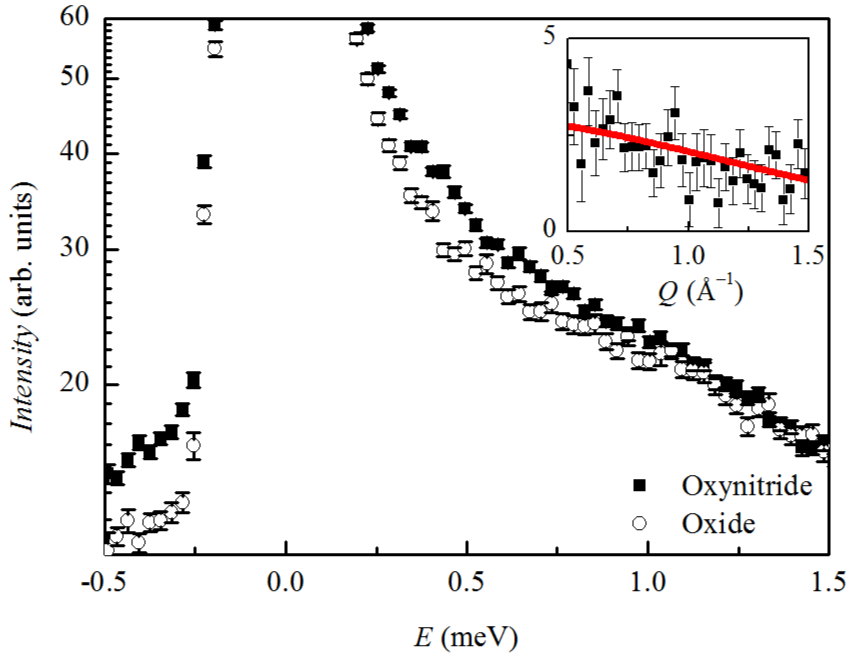}
\caption{The energy cuts of the oxide and oxynitride neutron scattering spectra at $1.5$ K, obtained by integrating over a range of $Q=[0.5,2.0]$ {\AA}$^{-1}$. The inset shows the $Q$-dependence of the background subtracted inelastic scattering from the oxynitride sample at the same temperature, to which an $F(Q)^2$ form was fitted (red solid line).}
\label{Figure4}
\end{figure}

Fig.{\ref{Figure4}} directly compares the inelastic scattering integrated over $Q=[0.5,2.0]$ {\AA}$^{-1}$ from the oxide and oxynitride samples at $1.5$ K. Given that the inelastic scattering from Lu$_2$Mo$_2$O$_7$ is strongly suppressed below the spin freezing transition, the scattering from the oxide at $1.5$ K is a good measure of the background. The oxynitride sample thus clearly shows an enhanced inelastic signal on either side of the elastic line. The persistence of inelastic scattering, particularly on the neutron energy gain side where the incident neutron destroys a spin fluctuation, directly demonstrates that the oxynitride remains strongly fluctuating at $1.5~{\sim}~|{\theta}|/100$ K. By subtracting the $1.5$ K oxide data from the oxynitride data and integrating in energy ($E=[0.2,1.5]$ meV) the $Q$-dependence of the inelastic scattering from the oxynitride is obtained (see Fig.{\ref{Figure4}}, inset). The form of this inelastic scattering reveals the nature of dynamical spin correlations within the oxynitride. The $Q$-dependence is well described by the magnetic form factor, $F(Q)^2$.  Taken together, the inelastic magnetic scattering from Lu$_2$Mo$_2$O$_5$N$_2$ strongly suggests a gapless QSL state at $1.5$ K, with few or no structured spatial correlations between the $S={\frac{1}{2}}$ spins.

Our study of Lu$_2$Mo$_2$O$_7$ and Lu$_2$Mo$_2$O$_5$N$_2$ present both interesting similarities and contrasts.  They share the same cubic pyrochlore structure and their magnetic interactions are antiferromagnetic and of similar strength.  Yet the $S=1$ Lu$_2$Mo$_2$O$_7$ displays an unconventional spin glass state below $T_f~{\sim}~16$ K, while the $S={\frac{1}{2}}$ Lu$_2$Mo$_2$O$_5$N$_2$ remains in a dynamic QSL state to temperatures less than $|{\theta}|/100$.  Even though structural disorder and magnetic frustration often lead to spin freezing, the oxynitride does {\it{not}} find a frozen ground state despite O/N disorder, unlike both Lu$_2$Mo$_2$O$_7$ and Y$_2$Mo$_2$O$_7$. These contrasts are clearly expressed in the low temperature magnetic heat capacities, where $C_{mag}~{\propto}~T^2$ for the oxides and $C_{mag}~{\propto}~T$ for the oxynitride. Quadratic temperature dependencies of heat capacity have been reported in a number of quasi-two-dimensional geometrically frustrated magnets with spin glass ground states, such as SCGO {\cite{ RamirezPRL1990}} and the jarosites {\cite{Jarosite_review}}.  For Y$_2$Mo$_2$O$_7$ {\cite{SilversteinPRB2014}} it was argued that this unusual temperature dependence arises from the coupling between spin and orbital degrees of freedom {\cite{MotomePRB2013}}.  Linear temperature dependencies have been observed in a number of QSL candidate materials  {\cite{Clark_DQVOF_2013}}, {\cite{deVries2008PRL}}, {\cite{YamashitaNatPhys2008}}, where it provides an upper limit for any gap, ${\Delta}$, in the  spin excitation spectrum.  In the case of the oxynitride, Lu$_2$Mo$_2$O$_5$N$_2$, this is ${\Delta}~{\sim}~0.5$ K $\sim~0.05$ meV.  Neutron spectroscopic data for the oxynitride (Figs.{\ref{Figure3}}(b) and {\ref{Figure4}}) set the upper limit for a gap at ${\sim}~0.1$ meV.

We finally turn to the relationship between the ground states of Lu$_2$Mo$_2$O$_7$ and  Y$_2$Mo$_2$O$_7$.  The ionic radius of Lu$^{3+}$ is ${\sim}~4~{\%}$ smaller than that of Y$^{3+}$ {\cite{ionic_radii}}, yet both oxides are``disorder free" spin glasses with similar $T_f$, $16$ K and $22$ K, respectively.  It is intriguing that a second pyrochlore antiferromagnet with a single magnetic site displays such a similar spin glass ground state.  If weak disorder is responsible for this state, it would imply that this disorder is very similar in the two materials - a conclusion which is not obvious based on their ionic radii.   Short range spin correlations within these two spin glass ground states are similar but distinct: Y$_2$Mo$_2$O$_7$ {\cite{GardnerPRL1999}} displays a broad magnetic diffuse elastic scattering peak centred at $0.4$ {\AA}$^{-1}$, as compared with the peak at $0.6$ {\AA}$^{-1}$ reported here for Lu$_2$Mo$_2$O$_7$. The lower $Q$ position of the peak of Y$_2$Mo$_2$O$_7$ indicates that its magnetic correlations have a longer ranged spatial extent. Indeed, those in Y$_2$Mo$_2$O$_7$ were found to form domains extending over an entire conventional unit cell, whereas the static magnetic correlations in Lu$_2$Mo$_2$O$_7$ are confined to a next-nearest-neighbour length scale, roughly half a unit cell.

To conclude, we observe a new realisation of an unconventional spin glass state in the $S=1$ molybdate oxide, Lu$_2$Mo$_2$O$_7$ with $T_f~{\sim}~16$ K.  This state is similar to that shown by its sister compound Y$_2$Mo$_2$O$_7$, although with shorter range static correlations peaked at $Q=0.6$ {\AA}$^{-1}$ in the magnetic scattering cross section.   The incorporation of the nitride anion into the oxide precursor tunes $S=1$ to $S={\frac{1}{2}}$ in Lu$_2$Mo$_2$O$_5$N$_2$. In contrast to the oxide, no evidence for any transition to a frozen or ordered state is observed to ${\sim}~0.5$ K and no elastic magnetic scattering is observed at temperatures as low as $1.5$ K.   The strength of the quantum fluctuations induced by the the reduction of $S$ in the oxynitride overcomes any competition with disorder in the selection of a ground state and gives rise to a QSL, with an upper bound on a gap in the spin excitation spectrum of ${\Delta}~{\sim}~0.05$ meV ${\sim}~0.5$ K or ${\frac{\Delta}{|\theta|}}~{\sim}~0.004$.   This study also demonstrates the potential of nitriding oxide precursors as a means of tuning the oxidation state on magnetic ions, and therefore the spin degree of freedom and strength of quantum fluctuations. This is a relatively unexplored route to exotic magnetic ground states, but one with the potential to be very rewarding.

\begin{acknowledgments}
Work at McMaster University was supported by NSERC of Canada. Work at the University of Edinburgh was supported by EPSRC and STFC. Research at Oak Ridge National Laboratory's Spallation Neutron
Source was supported by the Scientific User Facilities Division, Office of Basic Energy Sciences, U. S. Department of Energy. LC gratefully acknowledges useful discussions with J. R. Stewart, C. Stock and M. A. de Vries and support from the University of Edinburgh.  
\end{acknowledgments}

\section{Supplemental Material}
\section{Sample Preparation and Characterisation}

The oxynitride pyrochlore phase presented here was prepared by heat treatment of Lu$_2$Mo$_2$O$_7$ in ammonia gas at $600~^{\circ}$C for $24$ hours with a gas flow rate of $250$ cm$^3$min$^{-1}$. $3$ g of oxide was nitrided per run spread over a surface area of $15$ cm$^2$ in an alumina crucible  to give a single phase oxynitride product. Thermogravimetric and elemental analyses gave an average composition of Lu$_2$Mo$_2$O$_{4.8}$N$_{1.7}$.

\begin{figure}
\includegraphics[width=3.5in]{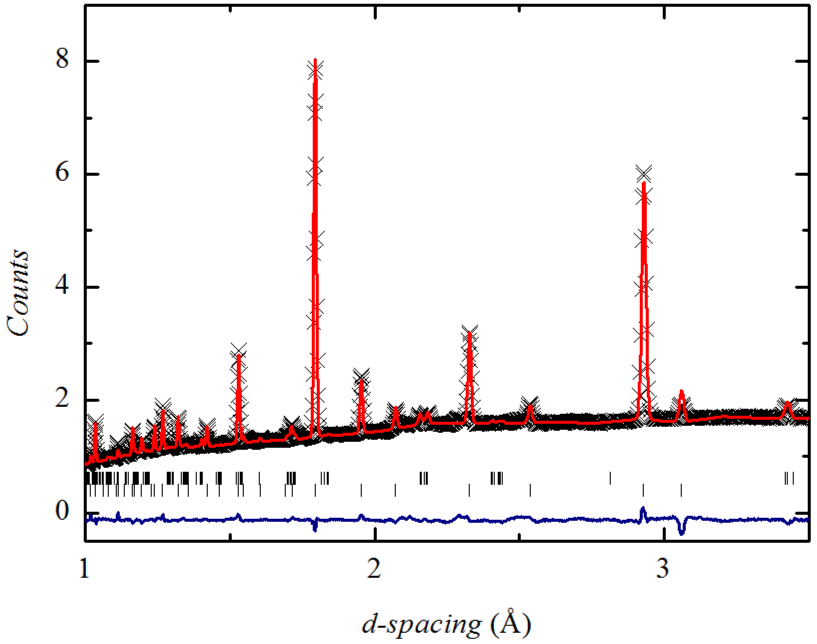}
\caption{Rietveld refinement of the cubic $Fd{\bar{3}}m$ model to 4 K HRPD data. Bottom tick marks show the reflections for
the oxynitride pyrochlore phase and top ticks mark a MoO$_2$ impurity phase (${{\sim}4~{\%}}$) that was present in the oxide precursor.}
\label{FigureS1}
\end{figure}

Powder neutron diffraction data were collected for the oxynitride sample on the High Resolution Powder Diffractometer (HRPD) at the ISIS spallation neutron source, Rutherford Appleton Laboratory, UK. $3$ g of sample was packed into a $6$ mm outer-diameter vanadium can and loaded into a He cryostat. Diffraction data were collected at $4$ K with a data collection time of $20$ hours. The cubic $Fd{\bar{3}}m$ model was refined to the data collected on the backscattering and $90~^{\circ}$ detector banks simultaneously using the GSAS software package {\cite{GSAS}}. The total anion content was constrained to the analytically determined values whilst allowing the oxide and nitride occupancies to refine over both anion sites. Isotropic thermal parameters were constrained together for cations and anions in order to minimise correlation effects. A plot of the fit to the $90^{\circ}$ data set is shown in Fig.{\ref{FigureS1}} and the results of the refinement are summarised in Table I. 

\begin{table}[htbp]
\caption{Refined atomic coordinates and occupancies for Lu$_2$Mo$_2$O$_{4.8}$N$_{1.7}$ ($a=10.1428(2)$ {\AA}). Isotropic thermal parameters were $0.0337(6)$ {\AA}$^2$ for metal cations and $0.0398(6)$ {\AA}$^2$ for anion sites. Total $R_{wp}=2.21~{\%}$, ${\chi}^2=14.58$ for $64$ variables.}
\begin{tabular}{cccccc}
\toprule
\ Atom & Site & $x$ & $y$ & $z$ & Occupancy \\
\colrule
Lu & $16d$ & $\frac{1}{2}$ & $\frac{1}{2}$ & $\frac{1}{2}$ & $1.0$ \\
Mo & $16c$ & $0$ & $0$ & $0$ & $1.0$ \\
O/N & $48f$ & $0.3477(1)$ & $\frac{1}{8}$ & $\frac{1}{8}$ & $0.663(2)$/$0.257$ \\
O$^{'}$/N$^{'}$ & $8b$ & $\frac{3}{8}$ & $\frac{3}{8}$ & $\frac{3}{8}$ & $0.831$/$0.169$ \\
\botrule
\end{tabular}
\end{table}

\begin{figure}
\includegraphics[width=3.5in]{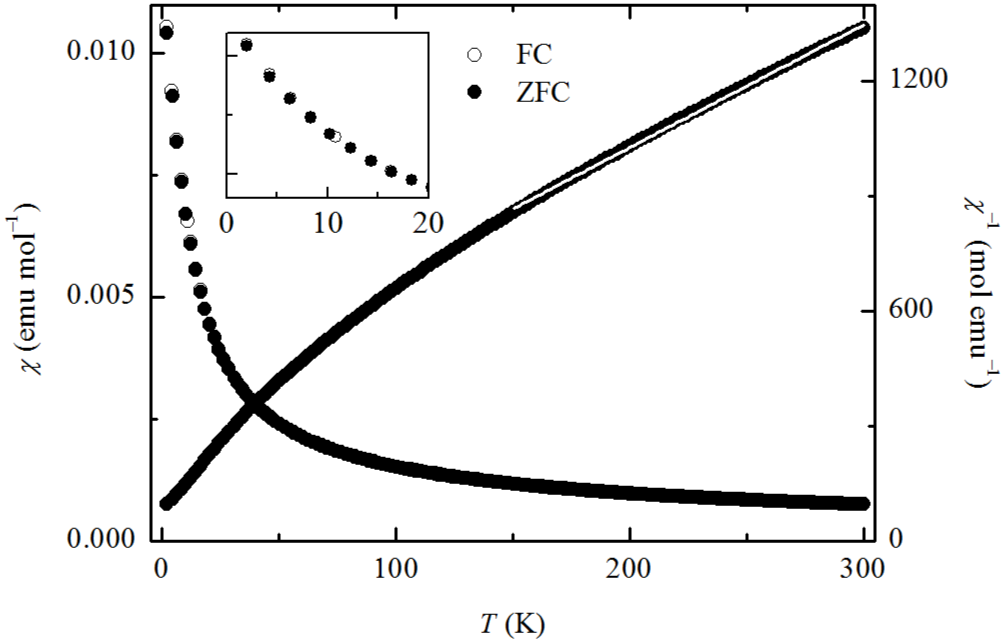}
\caption{The magnetic and inverse susceptibility data of Lu$_2$Mo$_2$O$_{5}$N$_{2}$ with a Curie-Weiss fit to the high temperature inverse susceptibility shown by the white solid line. The inset shows the FC and ZFC magnetic susceptibilities in the low temperature region.}
\label{FigureS2}
\end{figure}

The DC magnetic susceptibility of the oxynitride was measured in an applied field of 1 T from 2 K to 300 K in a zero field cooled (ZFC) field cooled (FC) cycle in a Quantum Design SQUID magnetometer. The magnetic and inverse susceptibilities of the oxynitride are shown in Fig.{\ref{FigureS2}}. The effective magnetic moment and Weiss constant extracted from the Curie Weiss fit depend on the temperature range over which the fit is applied. A similar temperature dependence is observed for the Curie Weiss fit to the inverse magnetic susceptibility of Lu$_2$Mo$_2$O$_7$ {\cite{ClarkJSSC2013}}. These results are summarised in Table II. There is no evidence of the spin freezing transition in the magnetic susceptibility that is observed for the oxide precursor at $16$ K {\cite{ClarkJSSC2013}}.

\begin{table}[htbp]
\caption{Results from the Curie Weiss fit to magnetic susceptibilities of the oxide and oxynitride pyrochlores.}
\begin{tabular}{cccc}
\toprule
\ Sample & Fit region / K & $-{\theta}$ / K & ${\mu}_{eff}$ / ${\mu}_B$ \\
\colrule
~ & $150 - 300$ & $158(1)$ & $1.89(1)$ \\
Lu$_2$Mo$_2$O$_7$ & $200 - 300$ & $171(1)$ & $1.92(1)$ \\
~ & $250 - 300$ & $184(2)$ & $1.85(1)$ \\
\colrule
~ & $150 - 300$ & $121(1)$ & $1.11(1)$ \\
Lu$_2$Mo$_2$O$_5$N$_2$ & $200 - 300$ & $135(1)$ & $1.13(1)$ \\
~ & $250 - 300$ & $152(2)$ & $1.16(1)$ \\
\botrule
\end{tabular}
\end{table}

\section{D$7$ Measurements}

\begin{figure}
\includegraphics[width=3.5in]{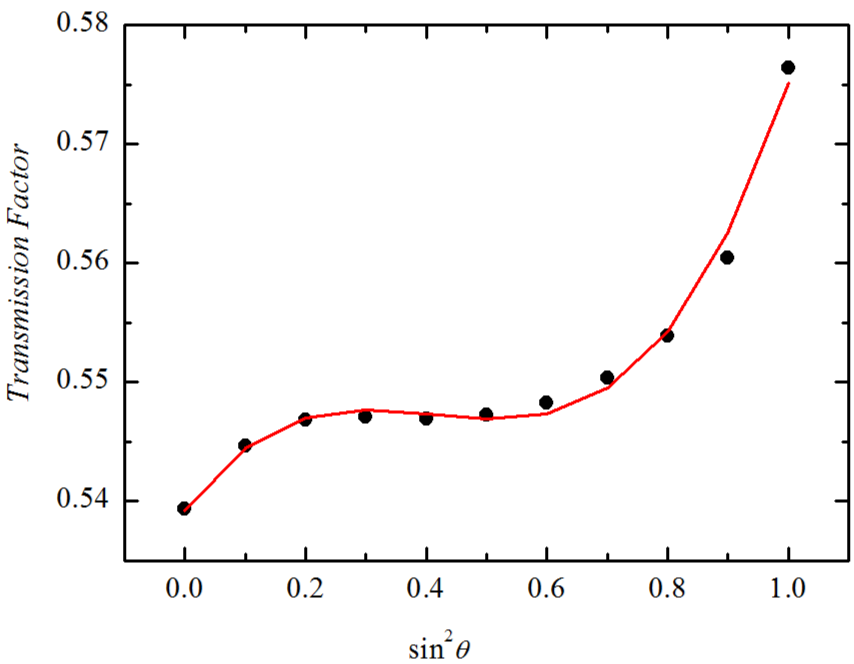}
\caption{The transmission factor for an annular cylindrical powder sample as a function of scattering angle taken from ref. {\cite{Schmitt1998}}, which was fitted by a third order polynomial function in ${\sin}^2{\theta}$  (red solid line).}
\label{FigureS3}
\end{figure}

\begin{figure}
\includegraphics[width=3.5in]{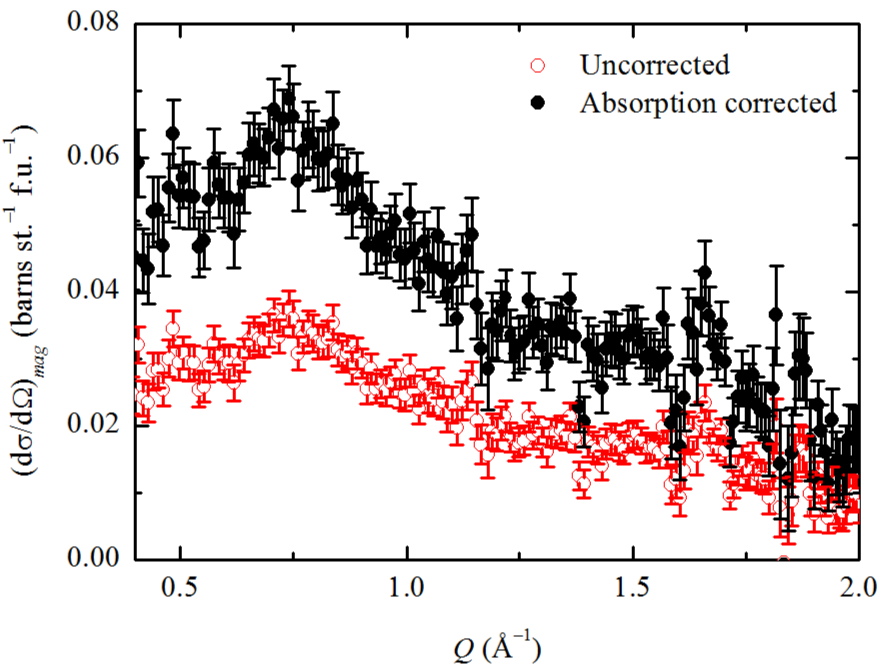}
\caption{The magnetic scattering cross section of Lu$_2$Mo$_2$O$_7$ measured at 1.5 K with an incident neutron wavelength ${\lambda}=4.8$ {\AA} before and after the angular dependent absorption correction for an annular cylindrical sample geometry.}
\label{FigureS4}
\end{figure}

The oxide and oxynitride samples, both with a mass of $18$ g, were packed individually into $2$ cm outer-diameter  aluminium cans with an annular geometry, in order to minimise beam attenuation from the samples. Calibration measurements of quartz, cadmium and vanadium standards allowed for the determination of the magnetic scattering cross sections of the samples in absolute units of barns st.$^{-1}$ per formula unit.  The transmission of the oxide sample at ${\lambda}=4.8$ {\AA} was determined by measuring the monitor counts of the direct beam through the sample ($C_S$) with the monitor counts of the beam through the empty can ($C_E$), normalised by the counts measured for a cadmium standard ($C_{Cd}$), which account for any neutrons that do not pass through the sample. Each of these monitor count measurements were taken for a total of $120$ s, such that the sample transmission is given by,
\begin{equation}
T = {\frac{C_{S} - C_{Cd}}{C_{E} - C_{Cd}}} = {\frac{5015-70}{10224 - 70}} = 0.49
\end{equation}
which reveals a significant attenuation of the neutron beam. As a result, the magnetic scattering cross sections of the oxide and oxynitride have been corrected for absorption. For a powder sample in an annular cylindrical geometry the attenuation of the incident neutron depends on two factors, ${\mu}R$ and ${\rho}$, where ${\mu}$ is the linear attenuation coefficient, $R$ is the outer radius of the annulus and $\rho$ is the ratio between the radii of the inner and outer cylinders in the annulus. For Lu$_2$Mo$_2$O$_7$, ${\mu}~{\sim}~0.8$ due to the large absorption cross section of lutetium {\cite{Scattering_lengths}}, and with inner and outer annulus radii of approximately $0.5$ cm and $1.0$ cm, respectively, ${\mu}R~{\sim}~0.8$ and ${\rho}~{\sim}~0.5$. The transmission factors for an annular cylindrical sample have been calculated elsewhere {\cite{Schmitt1998}} and are plotted in Fig.{\ref{FigureS3}} as a  function of ${\sin}^{2}{\theta}$ for ${\mu}R$ and ${\rho}$ values consistent with our experiment. The angular dependence of the transmission factor follows a third order polynomial function in ${\sin}^{2}{\theta}$ {\cite{Dwiggins1975}}. A fit of this form to the calculated sample transmission (Fig.{\ref{FigureS3}}) was performed to extract this angular dependence and was used to correct the data as a function of $Q$. A comparison of the magnetic scattering cross section of Lu$_2$Mo$_2$O$_7$ before and after the absorption correction is shown in Fig.{\ref{FigureS4}}. 

\section{CNCS Measurements}

\begin{figure}
\includegraphics[width=3.5in]{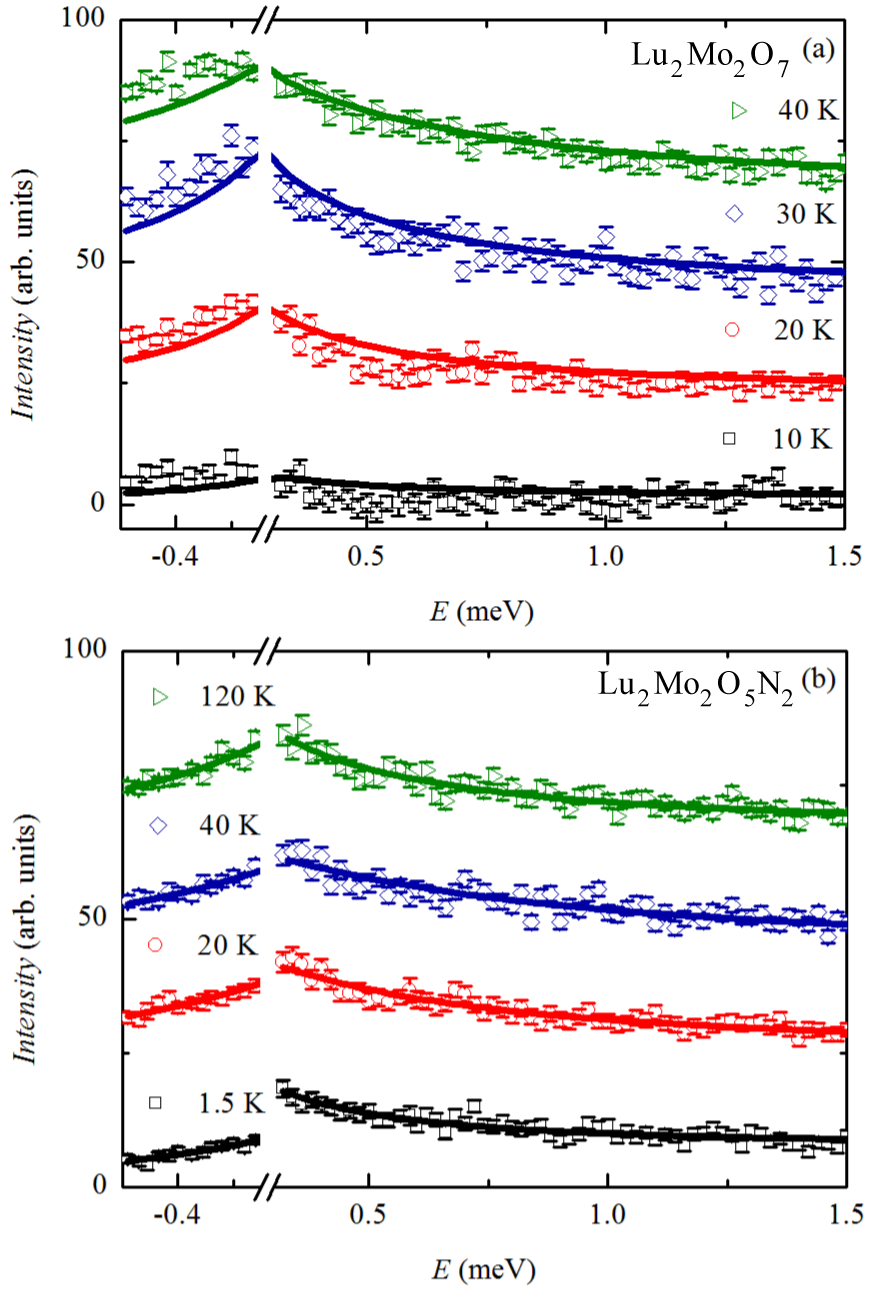}
\caption{$E$-cuts ($Q=[0.5,2.0]$ {\AA}$^{-1}$) for the (a) oxide and (b) oxynitride taken from background subtracted CNCS data at various temperatures. The solid lines are the fits to the inelastic scattering described in the manuscript.}
\label{FigureS5}
\end{figure}

For the collection of inelastic neutron spectra on the Cold Neutron Chopper Spectrometer (CNCS) the same $18$ g samples used in the D$7$ experiment were loaded into annular aluminium sample cans, in the same fashion that was employed in the diffuse scattering experiment. Individual cuts in $Q$ and $E$ were taken using the DAVE software package {\cite{DAVE}}. $E$-cuts were taken by integrating background corrected data over the $Q$-range [0.5,2.0] {\AA}$^{-1}$, which were modelled according to Equation $3$ given in the main manuscript. Fig.{\ref{FigureS5}} shows the fits to the data collected for the oxide and oxynitride at various temperatures. 

\bibliography{References}

\end{document}